\documentclass[12pt]{article}

\usepackage{amssymb}
\usepackage{a4wide}
\makeatletter
\def\fmslash{\@ifnextchar[{\fmsl@sh}{\fmsl@sh[0mu]}}
\def\fmsl@sh[#1]#2{%
  \mathchoice
    {\@fmsl@sh\displaystyle{#1}{#2}}%
    {\@fmsl@sh\textstyle{#1}{#2}}%
    {\@fmsl@sh\scriptstyle{#1}{#2}}%
    {\@fmsl@sh\scriptscriptstyle{#1}{#2}}}
\def\@fmsl@sh#1#2#3{\m@th\ooalign{$\hfil#1\mkern#2/\hfil$\crcr$#1#3$}}
\makeatother

\def\beq{\begin{eqnarray}}
\def\eeq{\end{eqnarray}}
\begin{document}
\begin{titlepage}
\begin{flushright}
SI-HEP-2006-16\\
November 8, 2006
\end{flushright}

\vfill

\begin{center}

{\Large\bf Minimal Flavour Violation and Beyond }\\[2cm]
{\large\bf  Th.~Feldmann and Th.~Mannel }\\[0.5cm]
{\it Theoretische Physik 1, Fachbereich Physik,
Universit\"at Siegen,\\ D-57068 Siegen, Germany }
\end{center}

\vfill

\begin{abstract}
\noindent

Starting from the effective-theory
framework for Minimal Flavour Violation, we give a
systematic definition of next-to-minimal (quark) flavour
violation in terms of a set of spurion fields exhibiting
a particular hierarchy with respect to a small
(Wolfenstein-like) parameter. A few illustrative
examples and their consequences for charged and
neutral decays with different quark chiralities
are worked out in some detail.
Our framework can be used as a model-independent classification 
scheme for the parameterization of flavour structure 
from physics beyond the Standard Model.

\end{abstract}

\vfill

\end{titlepage}

\clearpage




\section{Introduction}

\label{sec1}
%
%

%
Despite the enormous progress in the description of elementary
particle  interactions, the notion of flavour remains a mystery. In
the  standard model (SM) the flavour structure is parameterized by the
Yukawa couplings, which yield the masses  and the mixing angles of the
CKM matrix as physical parameters.
There are, at the moment, no convincing models for the observed
masses and mixing pattern, at least not at the quantitative level.  The
unification of forces in  Grand Unified Theories (GUTs, see e.g.\
\cite{Pati:1973uk,Georgi:1974sy,Fritzsch:1974nn})  mainly concerns the
gauge sector,  while flavour is still implemented by a triplication of
matter multiplets for the different fermion families, and the precise
mechanism  creating masses and mixings is parameterized in the symmetry
breaking sector.  In supersymmetric extensions of the SM, the soft
SUSY-breaking terms in the Lagrangian even add new sources of flavour
structure  (see e.g.\ \cite{Masiero:1997bv} and references therein for
a phenomenological discussion).

Due to the lack of a theory of flavour, we have no clear idea what
effects one may  expect beyond the parameterization encoded in the SM
Yukawa couplings.  With the next era of particle colliders in front of
us, and the hope to produce and detect new particles and interactions,
we also have to  improve the theoretical framework to discuss flavour
structure beyond the SM.
A well-known example is the concept of minimal flavour violation (MFV
\cite{D'Ambrosio:2002ex}, for an earlier introduction of the notion
see \cite{Ciuchini:1998xy,Buras:2000dm}),
which parametrizes new flavour effects by
the same two Yukawa  coupling matrices as they appear in the SM.  Up
to now all data in flavour physics, in particular from rare kaon and
$B$\/-meson decays, indicate that new-physics contributions to flavour
transitions are small.  Models with new physics at the TeV scale are
therefore favorably formulated within an MFV scenario. On the other
hand, MFV scenarios will shorten the  lever arm for flavour physics
experiments to discover and measure new physics in flavour
transitions, since -- except for the top quark --  all these
transitions involve  small mixing angles and/or Yukawa couplings.



The case for a super $B$ factory and the flavour-physics program at
the LHC lies in the hope that nature may be at some not too high scale
not  minimal flavour violating.  Again we do not have a compelling
theory for such a scenario, but we may  as well try to parameterize it.
In the present paper, we discuss a possible parameterization  in terms
of  additional spurion fields, which break the flavour symmetry in a
different  way as the two spurions associated with the Yukawa matrices
present already in the SM.
We will concentrate on the quark-flavour sector. Similar
considerations could also be performed for lepton-flavour transitions,
but will not be discussed in this paper. We will also stick to a
simple scenario with one Higgs doublet, but should keep in mind that
some flavour transitions can be enhanced by large $\tan\beta = \langle
H_1 \rangle/ \langle H_2 \rangle$ in 2-Higgs models. See
\cite{D'Ambrosio:2002ex} and \cite{Cirigliano:2005ck}
for discussions within MFV.

The paper is organized as follows. In the next section, we briefly
review the flavour structure following from the quark Yukawa couplings
to the Higgs field.  In section~\ref{sec4} we
summarize the flavour coefficients for quark transitions  within
MFV. Section~\ref{sec5} represents the main part of our paper, where
we give a possible -- model independent -- definition of
next-to-minimal flavour violating scenarios. For this purpose we
introduce additional spurion fields with different transformation
properties under the flavour group and a particular hierarchy with
respect to the Wolfenstein parameter $\lambda$.  Two illustrative
examples, where a new spurion --  coupling exclusively to right-handed
quarks -- appears,   are worked out in some more detail. We conclude
in section~\ref{sec6}.


\section{Quark-Flavours  in the Standard Model}

\label{sec2}

%

%
For quarks, the maximal flavour group which commutes with the 
gauge group of the SM is
\begin{equation} \label{FlavourGroup}
F =  SU(3)_{Q_L} \times  SU(3)_{U_R} \times  SU(3)_{D_R}
\end{equation}
where $Q_L$ denotes the weak doublets of left-handed quarks transforming as 
$(3,1,1)$, $U_R$ are the weak singlets of right-handed up-type quarks, 
transforming as $(1,3,1)$, and 
$D_R$ are the weak singlets of right-handed down-type quarks, transforming as 
$(1,1,3)$. The Higgs and the gauge fields of the SM transform as 
singlets under all factors of the flavour group (\ref{FlavourGroup}). 

The Yukawa couplings of the SM break the flavour symmetry 
(\ref{FlavourGroup}). This breaking can be described in terms of two spurion 
fields $Y_U$ and $Y_D$, 
where $Y_U$ is assumed to transform as $(3,\overline{3},1)$ and $Y_D$ as
$(3,1,\overline{3})$. The formally invariant terms with a single insertion 
of the spurions can be written as
\begin{eqnarray} 
- {\cal L}_{\rm yuk}
&=& 
\bar{Q}_L' H Y_D D_R' + 
\bar{Q}_L' H Y_U U_R' + {\rm h.c.} \, .
\label{eq:yuk1}
\end{eqnarray}
with the quark fields in the electro-weak basis written as
\begin{equation}
D_R' = 
\left( \begin{array}{c} 0 \\ d_R' \end{array} \right) \,,
\qquad 
U_R' = 
\left( \begin{array}{c} u_R' \\ 0 \end{array} \right) \,, 
\end{equation}
and the Higgs field introduced as a $2 \times 2$ matrix 
\begin{equation}
H = \frac{1}{\sqrt{2}} \left( \begin{array}{cc} 
                      \phi_0 + i \chi_0 & \sqrt{2} \phi_+ \\
                      \sqrt{2} \phi_-  & \phi_0 - i \chi \end{array} \right) \,.
\label{higgs}
\end{equation}
The VEV of the Higgs field is chosen to be $\langle \phi_0 \rangle = v \neq 0$ 
while the spurions $Y_U$ and $Y_D$ are ``frozen'' to the observed values   
of the Yukawa couplings. This leads to a mass term contained in 
(\ref{eq:yuk1}) and the mass eigenstates are obtained by diagonalizing 
the resulting mass matrices. 
This diagonalization procedure may be expressed by bi-unitary
transformations from the group $F$ given in (\ref{FlavourGroup}),\footnote{We
assume that the eigenvalues of the spurions are real and non-negative. 
This can always be achieved by an appropriate chiral rotation.}
\beq
  V_{u_L}^\dagger \, Y_U \, V_{u_R} &=& 
   \sqrt2 \,m_U^{\rm diag}/v \equiv \hat m_U \,, \\
  V_{d_L}^\dagger \, Y_D \, V_{d_R} &=& 
   \sqrt2 \, m_D^{\rm diag}/v \equiv \hat m_D \,,
\eeq
where $V_{u_L},V_{d_L} \in SU(3)_{Q_L}$,
$V_{u_R} \in SU(3)_{U_R}$ and $V_{d_R} \in SU(3)_{D_R}$.
This defines the quark fields $U,D$ in the mass eigenbasis
\begin{equation}
U_L^\prime = V_{u_L} \, U_L \,, \qquad U_R^\prime  = V_{u_R} U_R  \,, \qquad
D_L^\prime = V_{d_L} \, D_L \,, \qquad D_R^\prime  = V_{d_R} D_R  \,. 
\end{equation}
and the Yukawa interactions (\ref{eq:yuk1}) are expressed as
\beq 
-  {\cal L}_{\rm yuk}&=& 
\bar{Q}_L H \hat m_D D_R + \bar{Q}_L H \hat m_U U_R + {\rm h.c.} \, 
\label{eq:yuk2}
\eeq
In the mass eigenbasis the gauge sector of the SM reads
\beq
  {\cal L}_{\rm gauge} &=&
  \bar Q_L' \, i\fmslash D \, Q_L' + \bar U_R' \, i \fmslash D \, U_R'
      + \bar D_R' \, i \fmslash D \, D_R' \nonumber \\[0.2em]
  &=& \left(
\bar u_R \, i \fmslash D \, u_R + \bar d_R \, i \fmslash D \, d_R + (R
\to L) \right)
+ \left(\bar u_L \, V_{\rm CKM} \, i \fmslash D \, d_L + {\rm h.c.} \right) \,,
\eeq
where $i\fmslash D$ denotes the covariant derivative in the corresponding
representation of $SU(2)_L \otimes U(1)_Y$.
Here, the mismatch between $V_{u_L}$ and $V_{d_L}$ defines the
CKM matrix,
\beq
  V_{u_L}^\dagger  V_{d_L} & \equiv & V_{\rm CKM}
\eeq
and induces charged flavour transitions between $u_L$ and $d_L$. 

Thus the only observable flavour-violating effects in the SM 
(as well as in all  minimal flavour violating scenarios, to be
discussed below) are the
different quark masses and the relative rotation $V_{\rm CKM}$ 
between the two eigenbases defined by $V_{u_L}$ and $V_{d_L}$ in
which $Y_U$ and $Y_D$ are diagonal. Notice that the rotations $V_{u_R}$ and
$V_{d_R}$ are not observable in the SM.

%
\subsection{The Role of Custodial $SU(2)$ in Flavour Physics} 

\label{sec3}

%
%

It is often argued that the solution of the flavour problem will happen 
at some very high scale, possibly even the Planck scale. 
However, there is a symmetry connecting the flavour mixing and some properties 
of the mass spectrum with the gauge structure. 
This ``custodial'' symmetry \cite{custodial,custodial1,custodial2} 
is  an exact symmetry of the Higgs sector but is broken by the Yukawa couplings
and the fact that only one generator of the right handed symmetry is gauged, 
yielding the weak hypercharge.

More precisely, the Higgs sector of the SM has a
chiral $SU(2)_L \times SU(2)_R$ symmetry, under which the quark and
Higgs fields transform as 
$$
  Q_L \sim (2,1) \,, \qquad Q_R \sim (1,2) \,, \qquad H \sim (2, 2) \,.
$$
It is broken down to the custodial $SU(2)_{L+R}$
by the Higgs VEV $\langle \phi_0 \rangle \neq 0$. 
Under the remaining symmetry the three goldstone modes 
of the Higgs field in (\ref{higgs}) transform 
as a triplet, while the 
left- and right-handed up- and down-quarks form a doublet each. 
In the SM, custodial $SU(2)$ is explicitly broken by the gauge 
interactions and by the Yukawa couplings.

In case we enforce custodial $SU(2)$ as an additional symmetry, i.e.\ we 
assume that the flavour group commutes with the chiral symmetry 
$SU(2)_L \times SU(2)_R$, the flavour group to be considered
would reduce to
\begin{equation} \label{FlavourCust}
F_C =  SU(3)_{Q_L} \times  SU(3)_{U_R+D_R}
\end{equation}
since the right handed up and down quarks form a doublet under 
$SU(2)_{L+R}$.
For the Yukawa couplings of the left- and right-handed quarks this 
has the consequence
that there is only a single spurion field $Y_C$ transforming 
as $(3,\bar{3})$ under (\ref{FlavourCust}). Furthermore, making use of the 
freedom implied by (\ref{FlavourCust}) we may diagonalize $Y_C$ 
with the {\em same}\/ transformation for up and down quarks. 
Therefore, the presence of an exact custodial $SU(2)$ symmetry excludes 
the possibility of flavour mixing, and would imply
a degeneracy between the up and the down quark in each family.

In many GUTs (for instance in $SO(10)$) the right-handed up- and 
down-quarks of one family are assigned to the same multiplet 
of the gauge group. Thus custodial $SU(2)$ is a subgroup of the 
GUT gauge group, and the possible flavour group
collapses to (\ref{FlavourCust}),
in which case the possible Yukawa couplings can be
made diagonal and hence family mixing is absent.\footnote{For a
recent discussion of MFV within $SU(5)$ GUT, see \cite{Grinstein:2006cg}.}
This also means that the
origin of flavour mixing should be at or below the GUT scale, and related to
the scale where the breaking of custodial $SU(2)$ occurs.

%
\section{Minimal Flavour Violating New Physics}

\label{sec4}

%
%

%
\begin{table}[b!ht]
\begin{center}
\caption{\label{tab:MFV} \small
 Minimal number of spurion insertions to generate 
 flavour transitions between left- and right-handed up- and
 down quarks.}
\vspace{0.3em}
\footnotesize
\begin{tabular}{l || c | c | c | c}
\hline\hline 
           & $U_L$ & $U_R$  & $D_L$ & $D_R$ \\
\hline
$\bar U_L$ & $V_{u_L}^\dagger Y_D Y_D^\dagger V_{u_L}$ &
             $V_{u_L}^\dagger Y_D Y_D^\dagger Y_U V_{u_R}$ &
             $V_{u_L}^\dagger V_{d_L}$ &
             $V_{u_L}^\dagger Y_D V_{d_R}$ 
\\
           & ${} = V_{\rm CKM}  \hat m_D^2  V_{\rm CKM}^\dagger$ &
             ${} = V_{\rm CKM}  \hat m_D^2  V_{\rm CKM}^\dagger \hat m_U$ &
             ${} = V_{\rm CKM} $ &
             ${} = V_{\rm CKM}  \hat m_D  V_{\rm CKM}^\dagger$ 
\\
\hline
$\bar U_R$ & h.c. & 
             $V_{u_R}^\dagger Y_U^\dagger Y_D Y_D^\dagger Y_U V_{u_R}$ &
             $V_{u_R}^\dagger Y_U^\dagger V_{d_L}$ &
             $V_{u_R}^\dagger Y_U^\dagger Y_D V_{d_R}$ 
\\
           & &
             ${} = \hat m_U V_{\rm CKM}  \hat m_D^2  V_{\rm CKM}^\dagger \hat m_U$ &
             ${} = \hat m_U V_{\rm CKM} $ &
             ${} = \hat m_U V_{\rm CKM}  \hat m_D  V_{\rm CKM}^\dagger$ 
 \\
\hline
$\bar D_L$ & h.c. & 
             h.c. & 
             $V_{d_L}^\dagger Y_U Y_U^\dagger V_{d_L}$ &
             $V_{d_L}^\dagger Y_U Y_U^\dagger Y_D V_{d_R}$ 
\\
           & & 
             &
             ${} = V_{\rm CKM}^\dagger \hat m_U^2 V_{\rm CKM} $ &
             ${} =  V_{\rm CKM}^\dagger \hat m_U^2 V_{\rm CKM} \hat m_D$
 \\
\hline
$\bar D_R$ & h.c. & 
             h.c. & 
             h.c. &
             $V_{d_R}^\dagger Y_D^\dagger Y_U Y_U^\dagger Y_D V_{d_R}$ 
\\
           & & 
             &
             &
             ${} =\hat m_D V_{\rm CKM}^\dagger \hat m_U^2 V_{\rm CKM} \hat m_D$
\\ 
\hline\hline
\end{tabular}
\end{center}
\end{table}

The SM Lagrangian consists of all possible dimension-4 operators,
and the effect of switching to mass eigenstates is the flavour mixing  that
appears in the charged currents.  Using an effective field theory picture at
the electroweak scale $\mu \sim M_W$, possible new physics effects (arising
from some high scale $\Lambda \gg M_W$) can be parameterized by
higher-dimensional operators.  Due to the 
$SU(3)_c \times SU(2)_L \times U(1)_Y$
gauge symmetry the lowest  possible dimension for new operators involving
quarks is six,  and so a generic parameterization of new physics  in this
picture involves all possible dimension-six operators. While this concept is
quite successful in the gauge sector,  it involves too many parameters to be
useful  in the flavour sector.

It has been widely advertised to use the assumption of minimal flavour
violation  in order to reduce the number of possible parameters. Qualitatively
this means that  also in the new physics sector only the quark masses and
the CKM matrix are assumed to appear. 
A clear formulation of this concept has been given in
\cite{D'Ambrosio:2002ex},
and we shall use this approach here as well. 

Defining  MFV in the sense of \cite{D'Ambrosio:2002ex}, we have to look at   
insertions of the spurions $Y_U$ and $Y_D$ between quark fields, which
are consistent with the flavour group $F$. 
A complete list of the minimal number of insertions necessary to generate
flavour transitions between left- and right-handed up- and down-quark 
fields is given in Table~\ref{tab:MFV}. This includes the trivial case, 
i.e.\ no insertions at all for charged left-handed decays, leaving
the CKM matrix as in the SM. On the other hand, for right-handed FCNC
we need at least four spurion insertions. As a general rule, right-handed
decays in MFV involve an additional quark-mass factor per right-handed
field, and FCNC always involve at least two CKM elements. A special 
case of MFV is the weak effective Hamiltonian in the 
SM \cite{Buchalla:1995vs}, where the generic flavour
structures in Table~\ref{tab:MFV} are realized via box and penguin
diagrams.

An important point to notice here is that the predictive
power, following from the MFV assumption, is related to the fact that
most of the flavour structures in Table~\ref{tab:MFV} involve at
least one small CKM element and/or quark mass. Consequently,
the higher the number of spurion insertions, the smaller the 
corresponding coefficient. An exception to this rule are charged 
$t \to b$ transitions, where $m_t/v$ and $|V_{tb}|$ are of order one.
Therefore, new contributions to 
right-handed $t \to b$ transitions with ${\cal O}(1)$ 
flavour coefficients can occur even in MFV. 

From the possible MFV couplings for quark bilinears in Table~\ref{tab:MFV} one 
can easily construct the flavour couplings of all possible four-quark 
operators. The possible spin and colour structures are
constrained as usual by Lorentz and gauge symmetry, but their specification
is not relevant for the following discussion. In case of
rare semileptonic decays $q \to q' \ell^+\ell^-$, one would also
have to take into account the lepton-flavour sector. For simplicity, we
assume in the following, that the dominating effects come 
from new contributions to $q \to q'Z(\gamma)$ with subsequent SM couplings
of the gauge bosons to the lepton pair.


\section{Defining Non-Minimal Flavour Violation (nMFV)}

\label{sec5}


If MFV holds, the relative effects of new physics contributions to flavour 
transitions are as small as for flavour-diagonal processes. Actually, the
present experimental results for rare kaon and $B$\/-meson decays show no
evidence for inconsistencies with the SM, which can be taken as an indication
that -- if there is new physics around the TeV scale -- it is close to MFV.
On the other hand, if one allows for generic flavour transitions in 
higher-dimensional new-physics operators, one is forced to consider 
new-physics scales much larger than 1~TeV.

We may imagine an intermediate scenario, next-to-minimal flavour
violation (nMFV), where the size of the suppression factors for specific flavour 
transitions is somewhere between generic and minimal flavour violation.
In this chapter we are going to attempt a model-independent definition
of nMFV scenarios, using again a spurion analysis for the effective
theory at the electro-weak scale.\footnote{For an alternative approach, where
nMFV is defined by new physics coupling dominantly to the third generation,
see \cite{Agashe:2005hk}.}

Starting point are the quark bilinears in Table~\ref{tab:MFV} and 
their transformation under the flavour group $F$. There are
10 possible combinations of $3_{L,U,D}$ and $\bar 3_{L,U,D}$, namely
the flavour-singlet $(1,1,1)$ together with
\beq
 &&  (3,\bar 3, 1) \, \qquad (3, 1 , \bar 3) \, \qquad (1,3,\bar 3)  \qquad
    (\bar 3,3, 1) \, \qquad (\bar 3, 1 , 3) \, \qquad (1,\bar 3,3) \cr
 &&  (\bar 8,1,1) \, \qquad (1,8,1) \, \qquad (1,1,8) \,.
\eeq
The SM Yukawa couplings only involve $(3,\bar 3,1)$ and $(3,1,\bar 3)$ 
(and their conjugates), and therefore only the spurions $Y_U$ and $Y_D$
have to be considered. MFV is based on the assumption that $Y_U$ and $Y_D$
are sufficient to parameterize all relevant flavour transitions in new physics
operators. A possibility to define nMFV is to allow for one (or more)
additional elementary spurion fields from the following set,
\beq
&&  Y_R \sim (1,3,\bar 3) \,, \qquad \qquad \  Y_R^\dagger \sim (1,\bar 3,3) 
\nonumber \\[0.2em]
&&  Z_L=Z_L^\dagger \sim (8,1,1) \,, \qquad
  Z_U=Z_U^\dagger \sim (1,8,1) \,, \qquad
  Z_D=Z_D^\dagger \sim (1,1,8) \,.
\label{eq:new}
\eeq
In order achieve predictive power, we have again to require that
the elements of the new spurion fields show some hierarchy in terms
of a small parameter $\lambda'$ (similar, but not necessarily related 
to the Wolfenstein parameter $\lambda$). 

The new spurion fields and their combinations with
the MFV spurions $Y_U$ and $Y_D$ give new (independent) possibilities
to saturate the flavour structures in Table~\ref{tab:MFV}. In some
cases, one needs a {\em smaller}\/ number of spurion insertions than
in MFV, i.e.\ one generates potentially {\em larger}\/ flavour coefficients. 
The possibility to combine nMFV and MFV spurion fields 
constrains the allowed power-counting for the nMFV spurions.
For instance, the combination $Y_U Y_R \sim (3,1,\bar 3)$ transforms
as $Y_D$, and therefore it can also appear at the corresponding
place in the SM Yukawa term. In order to keep the SM power counting
for CKM angles and quark masses, we thus have to require that
\beq
  (Y_U Y_R)_{ij} \sim (\lambda^{(')})^{n_{ij}} 
 &\leq& (\lambda)^{m_{ij}} \sim (Y_D)_{ij} \qquad \mbox{etc.}
\label{ineq}
\eeq
for all $i,j$, where $n_{ij}$ and $m_{ij}$ are some integer
numbers specifying the power-counting in a given new-physics
model. If these inequalities hold, we can always absorb
the effects of nMFV spurions appearing in dim-4 operators 
into a redefinition of the MFV spurions $Y_U$ and $Y_D$.

To illustrate our idea, we will, in the following, consider an example,
where we include one nMFV spurion $Y_R$. In this case, we can express all 
possible flavour coefficients in terms of the quark masses, the CKM
matrix and a new complex matrix\footnote{In models with right-handed
gauge bosons $W'$, the matrix $R$ can be identified with the CKM matrix 
$V_{\rm CKM}'$ in the right-handed sector. In this case, $R^\dagger = R^{-1}$ 
is unitary. In the general nMFV scenario $R R^\dagger \neq 1$.}
$$
  R = V_{u_R}^\dagger Y_R V_{d_R} \,.
$$ 
For instance,
\beq
  V_{u_L}^\dagger Y_U Y_R V_{d_R} &=& 
  \hat m_U \, V_{u_R}^\dagger Y_R  V_{d_R} \equiv  \hat m_U \, R  \,,\\
  V_{d_L}^\dagger Y_U Y_R V_{d_R} &=& 
  V_{\rm CKM}^\dagger \, \hat m_U \, V_{u_R}^\dagger Y_R  V_{d_R} 
    \equiv  \hat  V_{\rm CKM}^\dagger \,\hat m_U \, R \,,\\
  V_{d_L}^\dagger Y_D Y_R^\dagger V_{u_R} &=& 
  \hat m_D \, V_{d_R}^\dagger Y_R^\dagger  V_{u_R} 
   \equiv  \hat m_D \, R^\dagger  \,,\\
  V_{u_L}^\dagger Y_D Y_R^\dagger V_{u_R} &=& 
  V_{\rm CKM} \, \hat m_D \, V_{d_R}^\dagger Y_R^\dagger  V_{u_R} 
   \equiv  V_{\rm CKM}  \, \hat m_D \, R^\dagger \,, \quad
\mbox{etc.}
\eeq
In particular, since $Y_R$ exclusively couples to right-handed quarks,
it can induce potentially large effects in right-handed flavour 
transitions which are suppressed by small Yukawa couplings in the SM.
Of course, the size of the effects depends crucially on the assumed
power-counting for the matrix elements $R_{ij}$.
In the next subsections, we specify two examples, where
the power counting for $R_{ij}$ is fixed either within a
simple Froggatt-Nielsen model, or assumed to be democratic (i.e.\
independent of $i$ and $j$).


\subsection{Example 1: \\ nMFV spurion $Y_R$ and Froggatt-Nielsen power-counting}


To illustrate the possible quantitative effects of nMFV flavour
structures we use a (minimal) Froggatt-Nielsen scenario (FN) 
\cite{Froggatt:1978nt} (see also \cite{Leurer:1992wg}). 
In this scenario the flavour transitions are due to interactions
with some scalar field which breaks a hypothetical $U(1)$ symmetry
at a high scale.
Different quark multiplets (in the weak eigenbasis)
are supposed to have different charges under that symmetry:
\beq
&&  Q_L^i: c + b_i \,,\qquad
    U_R^i: c - a_i^u \,, \qquad
    D_R^i: c - a_i^d  \,.
\eeq
The hierarchy of the Yukawa couplings then follows as
\beq
&&  (Y_U)_{ij} \sim \lambda^{|b_i +a_j|} 
  \,, \qquad  (j=u,c,t)
\cr
&&
    (Y_D)_{ij} \sim \lambda^{|b_i +a_j|}
  \,,  \qquad (j=d,s,b)
\eeq
where $\lambda$ is the ratio of the VEV of the new scalar field
and the new-physics scale, and is to be identified with
the Wolfenstein parameter. $SU(2)_L$ invariance requires
$b_u=b_d\equiv b_{u,d}$, $b_c=b_s\equiv b_{c,s}$, 
$b_t=b_b\equiv b_{t,b}$.
One further assumes
$a_i > 0$ and $b_i \geq 0$, together with
the ordering $a_u \geq a_c \geq a_t$,
                        $a_d \geq a_s \geq a_b$, and
                        $b_{u,d} \geq b_{c,s} \geq b_{t,b}$.
The eigenvalues of up- and down-quark mass matrices follow as 
$$m_i \sim \lambda^{b_i +a_i}\,.$$
The CKM elements scale as
$$
  (V_{\rm CKM})_{ij} \sim \lambda^{|b_i-b_j|} \,.
$$
The Wolfenstein counting for the CKM matrix thus fixes
the differences for the charges $b_i$,
$$
  b_{u,d}-b_{c,s} = 1 \,, \quad
  b_{c,s}-b_{t,b} = 2 \,, \quad
  b_{u,d}-b_{t,b} = 3 \,.
$$
Notice that the Wolfenstein counting for the quark masses can
independently be controlled by the parameters $a_i$. 
Two phenomenologically acceptable examples are listed 
in Table~\ref{tab:pc}, where
we fixed the unobservable $b_{t,b}=0$ for simplicity.

\begin{table}[t!]
\begin{center}
\caption{\label{tab:pc} \small Two examples for 
FN charges and the related Wolfenstein power-counting for quark masses.
For simplicity, we fixed $b_{t,b}=0$.}
\vspace{0.3em}
   \begin{tabular}{ccc|ccc|ccc||ccc|ccc}
\hline\hline
      $a_u$ & $a_c$ & $a_t$ & 
      $a_d$ & $a_s$ & $a_b$ & 
      $b_{u,d}$ & $b_{c,s}$ & $b_{t,b}$ & 
      $m_u$ & $m_c$ & $m_t$ & 
      $m_d$ & $m_s$ & $m_b$ \\
\hline\hline 
      5 & 2 & 0 &
      4 & 3 & 2 &
      3 & 2 & 0 &
      $\lambda^8$ & $\lambda^4$ & $\lambda^0$ &
      $\lambda^7$ & $\lambda^5$ & $\lambda^2$
\\
      3  & 1 & 0 &
      3  & 2 & 2 &
      3  & 2 & 0 &
      $\lambda^6$ & $\lambda^3$ & $\lambda^0$ &
      $\lambda^6$ & $\lambda^4$ & $\lambda^2$
\\
\hline \hline
  \end{tabular}
\end{center}
\end{table}

If we introduce other spurions with elementary transformations
under the flavour group, the power-counting is fixed by the FN charges,
too.
For $Y_R$, in particular, we obtain
$$
  (Y_R)_{ij} \sim \lambda^{|a_i - a_j|}  \qquad
  (i=u,c,t;\ j=d,s,b)
$$
For the first (second) example in Table~\ref{tab:pc}, the
power counting reads
\beq
  Y_R & \sim & \left( \matrix{ 
  \lambda^{1(0)} & \lambda^{2(1)} & \lambda^{3(1)} \cr
  \lambda^{2(2)} & \lambda^{1(1)} & \lambda^{0(1)} \cr
  \lambda^{4(3)} & \lambda^{3(2)} & \lambda^{2(2)}
   } \right) \,.
\label{eq:YRscale}
\eeq
Indeed, triangle inequalities between the FN charges guarantee that 
combinations of $Y_R$ with $Y_U$ or $Y_D$ do not lead to larger terms 
than those already present in the SM,
\beq
  (Y_D Y_R^\dagger)_{ij} \sim \lambda^{|b_i + a_{j'}|+|-a_j'+a_j|}
        &\leq & \lambda^{|b_i +a_j|} \sim (Y_U)_{ij}
\\[0.3em]
  (Y_U Y_R)_{ij} \sim \lambda^{|b_i + a_{i'}|+|-a_{i'}+a_j|}
        &\leq & \lambda^{|b_i +a_j|} \sim (Y_D)_{ij}
\eeq
On the other hand, the elements of $Y_R$ can be larger than the
corresponding flavour structures that one can build from $Y_U$ and
$Y_D$ in MFV,
\beq
  (Y_U^\dagger Y_D)_{ij} 
   \sim \lambda^{|a_i + b_k|+|-b_k - a_j|} & \leq &
     \lambda^{|a_i-a_j|} \sim (Y_R)_{ij}
\eeq
Below, we will systematically study the effect of $Y_R$ insertions
with FN power counting for charged and neutral flavour transitions 
with different chiralities.

\paragraph{Charged Decays}

\begin{table}[t!bh]
\begin{center} 
\caption{\label{tab:charged} \small
Charged currents in MFV and nMFV. The global suppression factor for
new-physics contributions to dim-6 operators 
is $v^2/\Lambda_{\rm NP}^2$. The quoted relative suppression
(or enhancement)
factors are to be understood with respect to the leading 
left-handed SM transitions. The Wolfenstein power-counting
refers to the Froggatt-Nielsen scenario 
with two alternatives for the power-counting of quark masses 
in the first (second) row of Table~\ref{tab:pc}. }
\vspace{0.3em}
\begin{tabular}{ c|| c || c | c || c | c }
\hline \hline
decay & SM & MFV & rel.~factor & nMFV & rel.~factor
\\
\hline
$\bar U_L D_L$ & $|V_{UD}|$ & $|V_{UD}|$ & 1 & - & - \\
$\bar U_L D_R$ &   & $\hat m_D |V_{UD}|$ & 
                     $\lambda^{|a_D+b_D|}$ & $Y_U Y_R$ &  
                       $\lambda^{|a_U + b_U|+|a_U-a_D| - |b_U-b_D|}$  \\
$\bar U_R D_L$ &   &$\hat m_U |V_{UD}|$ & $\lambda^{a_U+b_U}$ 
& $Y_R Y_D^\dagger$ & $\lambda^{|a_D + b_D|+|a_U-a_D| - |b_U-b_D|}$   \\
$\bar U_R D_R$ &   &$\hat m_U \hat m_D |V_{UD}|$ & $\lambda^{a_U+b_U +a_D+b_D}$ & $Y_R$
& $\lambda^{|a_U-a_D| - |b_U-b_D|}$ \\
\hline
\hline
\end{tabular}
\end{center}
\end{table}

The Wolfenstein power-counting for charged flavour transitions
with different chiralities for MFV and nMFV (with FN power-counting
for the spurion $Y_R$) are summarized in Table~\ref{tab:charged}.
Here we used that (in the mass basis) 
the entries of the matrix $R$ have the same power counting as those 
of $Y_R$, since in FN the rotation matrices
$V_{u_R}$ and $V_{d_R}$ are unity up to order $\lambda$ effects.
The interesting quantity is the suppression/enhancement factor,
coming from the new possible flavour structures involving the
spurion $Y_R$, relative to the leading left-handed tree-level
transition in the SM. From the last column in Table~\ref{tab:charged}
we find
\beq
  \bar U_L^i D_R^j &:& \frac{v^2}{\Lambda_{\rm NP}^2} 
       \left( \matrix{ 
    \lambda^{9(6)} & \lambda^{9(6)} & \lambda^{8(4)} \cr
    \lambda^{5(4)} & \lambda^{5(4)} & \lambda^{2(2)} \cr
    \lambda^{1(0)} & \lambda^{1(0)} & \lambda^{2} } \right)_{ij} \,,
\\[0.2em]
  \bar U_R^i D_L^j &:& \frac{v^2}{\Lambda_{\rm NP}^2} 
       \left( \matrix{ 
    \lambda^{8(6)} & \lambda^{6(4)} & \lambda^{2(0)} \cr
    \lambda^{8(7)} & \lambda^{6(5)} & \lambda^{0(1)} \cr
    \lambda^{8(6)} & \lambda^{6(4)} & \lambda^{4} } \right)_{ij} \,,
\\[0.2em]
  \bar U_R^i D_R^j &:& \frac{v^2}{\Lambda_{\rm NP}^2} 
       \left( \matrix{ 
    \lambda^{1(0)} & \lambda^{1(0)} & \lambda^{0(-2)} \cr
    \lambda^{1(1)} & \lambda^{1(1)} & \lambda^{-2(-1)} \cr
    \lambda^{1(0)} & \lambda^{1(0)} & \lambda^{2} } \right)_{ij} \,,
\eeq
where the exponents refer to the power-counting for quark masses
in the first (second) row of Table~\ref{tab:pc}. For purely
left-handed transitions the spurion $Y_R$ can only appear in
combinations like $Y_U Y_R Y_D^\dagger$ which are always smaller
than $V_{\rm CKM}$ due to the triangle inequalities holding in
FN.

\paragraph{FCNCs involving down-quarks}

In Table~\ref{tab:neutral_d} we summarize the spurion combinations
contributing to FCNCs with $d$\/-quarks in MFV and nMFV (with
Wolfenstein power-counting for $Y_R$ from FN).
Again, for purely left-handed FCNC, insertions of $Y_R$ cannot lead
to larger effects than in MFV.
For transitions with one or two right-handed down quarks, we find
for the suppression/enhance\-ment factors relative to the SM case,
\beq
  \bar D_L^i D_R^j &:& \frac{16 \pi^2 v^2}{\Lambda_{\rm NP}^2} 
       \left( \matrix{ 
            - & \lambda^{1(0)} & \lambda^{2} \cr
    \lambda^{1(0)} & - & \lambda^{2} \cr
    \lambda^{1(0)} & \lambda^{1(0)} & - } \right)_{ij} \,,
\\[0.2em]
  \bar D_R^i D_R^j &:& \frac{16 \pi^2 v^2}{\Lambda_{\rm NP}^2} 
       \left( \matrix{ 
    - & \lambda^{-2(-4)} & \lambda^{-1(-2)} \cr
    \lambda^{-2(-4)} & - & \lambda^{-1(0)} \cr
    \lambda^{-1(-2)} & \lambda^{-1(0)} & - } \right)_{ij} \,.
\eeq
To derive the second column in Table~\ref{tab:neutral_d},
we have used that 
$$
V_{d_L}^\dagger Y_U Y_R V_{d_R} = V_{\rm CKM}^\dagger \hat m_U R \,,
$$
and $b_t = a_t = 0$.

\begin{table}[h!]
\begin{center} 
\caption{\label{tab:neutral_d} \small
FCNCs with down quarks in MFV and nMFV. 
The global suppression factor for
new-physics contributions to dim-6 operators 
is $16\pi^2 v^2/\Lambda_{\rm NP}^2$. 
The quoted relative suppression (or enhancement)
factors are to be understood with respect to the leading 
left-handed (loop-induced) SM transitions. 
The Wolfenstein power-counting refers to the Froggatt-Nielsen scenario 
with two alternatives for the power-counting of quark masses 
in the first (second) row of Table~\ref{tab:pc}.}
\vspace{0.3em}
\begin{tabular}{ c|| c | c || c | c }
\hline \hline
decay & \multicolumn{2}{c||}{SM + MFV} & nMFV & rel.~factor
\\
\hline
$\bar D_L D_L'$ & $Y_U Y_U^\dagger$ & $\hat m_t^2 |V_{tD'}V_{tD}^*|$ & - & -
\\
$\bar D_L D_R'$ & $Y_U Y_U^\dagger Y_D$ & 
                  $ \hat m_D' \hat m_t^2 |V_{tD'}V_{tD}^*|$ & 
                  $Y_U Y_R$ & $\lambda^{|a_D'|-|b_D'|}$
\\
$\bar D_R D_R'$ & $Y_D^\dagger  Y_U Y_U^\dagger Y_D$ & 
                  $ \hat m_D \hat m_D' \hat m_t^2 |V_{tD'}V_{tD}^*|$ & 
                  $Y_R^\dagger Y_R$ & 
                  ${\rm max}_u 
          \left[\lambda^{|a_D-a_u|+|a_D'-a_u|-|b_D|-|b_D'|}\right]$
\\
\hline
\hline
\end{tabular}
\end{center}
\end{table}

\begin{table}[h!]
\begin{center} 
\caption{\label{tab:neutral_u} \small
FCNCs with up quarks in MFV and nMFV. 
The global suppression factor for
new-physics contributions to dim-6 operators 
is $16\pi^2 v^2/\Lambda_{\rm NP}^2$. 
The quoted relative suppression (or enhancement)
factors are to be understood with respect to the leading 
left-handed (loop-induced) SM transitions. 
The Wolfenstein power-counting refers to the Froggatt-Nielsen scenario 
with two alternatives for the power-counting of quark masses 
in the first (second) row of Table~\ref{tab:pc}.}
\vspace{0.3em}
\begin{tabular}{ c|| c | c || c | c }
\hline \hline
decay & \multicolumn{2}{c||}{ SM + MFV} & nMFV & rel.~factor
\\
\hline
$\bar U_L U_L'$ & $Y_D Y_D^\dagger$ & $\hat m_b^2 |V_{Ub}V_{U'b}^*|$ & - & -
\\
$\bar U_L U_R'$ & $Y_D Y_D^\dagger Y_U$ & $ \hat m_U' \hat m_b^2 |V_{Ub}V_{U'b}^*|$ & 
                  $Y_D Y_R^\dagger$ &
  $\lambda^{|a_b-a_U'|-a_b-b_U'}$
\\
$\bar U_R U_R'$ & $Y_U^\dagger Y_D Y_D^\dagger Y_U$ &
                  $ \hat m_U \hat m_U' \hat m_b^2 |V_{Ub}V_{U'b}^*|$ & 
                  $Y_R Y_R^\dagger$ &
                  ${\rm max}_d 
          \left[\lambda^{|a_U-a_d|+|a_U'-a_d|-b_U-b_U'-2 a_b}\right]$
\\
\hline
\hline
\end{tabular}
\end{center}
\end{table}

\paragraph{FCNCs involving up-quarks}

In Table~\ref{tab:neutral_u} we summarize the spurion combinations
contributing to FCNCs with $u$\/-quarks in MFV and nMFV (with
Wolfenstein power-counting for $Y_R$ from FN).
Again, for purely left-handed FCNC, insertions of $Y_R$ cannot lead
to larger effects than in MFV.
For transitions with one or two right-handed up quarks, we find
for the suppression/enhancement factors relative to the SM case,
\beq
  \bar U_L^i U_R^j &:& \frac{16 \pi^2 v^2}{\Lambda_{\rm NP}^2} 
       \left( \matrix{ 
            - & \lambda^{-4(-3)} & \lambda^0 \cr
    \lambda^{-2(-4)} & - & \lambda^0 \cr
    \lambda^{-2(-4)} & \lambda^{-4(-3)} & - } \right)_{ij} \,,
\\[0.2em]
  \bar U_R^i U_R^j &:& \frac{16 \pi^2 v^2}{\Lambda_{\rm NP}^2} 
       \left( \matrix{ 
    - & \lambda^{-6(-7)} & \lambda^{-2(-4)} \cr
    \lambda^{-6(-7)} & - & \lambda^{-4(-3)} \cr
    \lambda^{-2(-4)} & \lambda^{-4(-3)} & - } \right)_{ij} \,.
\eeq


\subsection{Example 2: \\  nMFV spurion $Y_R$ with democratic power-counting}


The simple FN scenario in the previous section clearly leads to rather 
large effects in certain flavour transitions and therefore should not 
be considered as phenomenologically favorable. An alternative and 
complementary approach would be to assume a {\it democratic}
power counting for the new spurion fields (in the mass eigenbasis). 
Sticking again to
the scenario with one nMFV spurion $Y_R$, we consider the power-counting
\begin{eqnarray}
  R_{ij} &\sim& \lambda^{4(3)}
\label{YRdem}
\end{eqnarray}
which corresponds to the smallest entry in (\ref{eq:YRscale}), where we consider
again the Wolfenstein-scaling for quarks as in Table~\ref{tab:pc}. 
The relative suppression/enhancement factors with respect to the leading SM contributions
in this case are as follows.

\paragraph{Charged Decays}

\beq
  \bar U_L^i D_R^j &:& \frac{v^2}{\Lambda_{\rm NP}^2} 
       \left( \matrix{ 
    \lambda^{12(9)} & \lambda^{11(8)} & \lambda^{9(6)} \cr
    \lambda^{7(5)} & \lambda^{8(6)} & \lambda^{6(4)} \cr
    \lambda^{1(0)} & \lambda^{2(1)} & \lambda^{4(3)} } \right)_{ij} \,,
\\[0.2em]
  \bar U_R^i D_L^j &:& \frac{v^2}{\Lambda_{\rm NP}^2} 
       \left( \matrix{ 
    \lambda^{11(9)} & \lambda^{8(6)} & \lambda^{3(2)} \cr
    \lambda^{10(8)} & \lambda^{9(7)} & \lambda^{4(3)} \cr
    \lambda^{8(6)} & \lambda^{7(5)} & \lambda^{6(5)} } \right)_{ij} \,,
\\[0.2em]
  \bar U_R^i D_R^j &:& \frac{v^2}{\Lambda_{\rm NP}^2} 
       \left( \matrix{ 
    \lambda^{4(3)} & \lambda^{3(2)} & \lambda^{1(0)} \cr
    \lambda^{3(2)} & \lambda^{4(3)} & \lambda^{2(1)} \cr
    \lambda^{1(0)} & \lambda^{2(1)} & \lambda^{4(3)} } \right)_{ij} \,,
\eeq

\paragraph{FCNCs involving down-quarks}

\beq
  \bar D_L^i D_R^j &:& \frac{16 \pi^2 v^2}{\Lambda_{\rm NP}^2} 
       \left( \matrix{ 
            - & \lambda^{2(1)} & \lambda^{4(3)} \cr
    \lambda^{1(0)} & - & \lambda^{4(3)} \cr
    \lambda^{1(0)} & \lambda^{2(1)} & - } \right)_{ij} \,,
\\[0.2em]
  \bar D_R^i D_R^j &:& \frac{16 \pi^2 v^2}{\Lambda_{\rm NP}^2} 
       \left( \matrix{ 
    - & \lambda^{3(1)} & \lambda^{5(3)} \cr
    \lambda^{3(1)} & - & \lambda^{6(4)} \cr
    \lambda^{5(3)} & \lambda^{6(4)} & - } \right)_{ij} \,.
\eeq

\paragraph{FCNCs involving up-quarks}

\beq
  \bar U_L^i U_R^j &:& \frac{16 \pi^2 v^2}{\Lambda_{\rm NP}^2} 
       \left( \matrix{ 
            - & \lambda^{0(-1)} & \lambda^{2(1)} \cr
    \lambda^{-1(-2)} & - & \lambda^{2(1)} \cr
    \lambda^{-1(-2)} & \lambda^{0(-1)} & - } \right)_{ij} \,,
\\[0.2em]
  \bar U_R^i U_R^j &:& \frac{16 \pi^2 v^2}{\Lambda_{\rm NP}^2} 
       \left( \matrix{ 
    - & \lambda^{-1(-3)} & \lambda^{1(-1)} \cr
    \lambda^{-1(-3)} & - & \lambda^{2(0)} \cr
    \lambda^{1(-1)} & \lambda^{2(0)} & - } \right)_{ij} \,.
\eeq


\subsection{Phenomenological implications}


As already stated in the introduction, the concept of MFV provides
a natural explanation for the present success of the SM in reproducing
the flavour observables in the CKM analysis, despite the possible
existence of new physics at or slightly below the TeV scale.
Within MFV the phenomenological
determination of quantities like $|V_{ub}|$ from $b \to u\ell \nu$,
$|V_{td}/V_{ts}|$ from $\Delta M_{B_d}/\Delta M_{B_s}$ or
$\Gamma[b\to d\gamma]/\Gamma[b\to s\gamma]$,
and $\sin2\beta$ from $|a_{J/\psi K}^{\rm CP}|$ is insensitive to new 
physics effects, even in 2-Higgs scenarios with large-$\tan\beta$ 
\cite{D'Ambrosio:2002ex}.

%

However, for the same reason, it will be difficult to really establish
minimally flavour-violating new physics in flavour transitions. 
On the one hand, one has to identify
small deviations from the SM. On the other hand, 
in order to exclude nMFV, one has to show that all flavour transitions 
are indeed driven by the CKM and mass factors as predicted by the analysis of 
\cite{D'Ambrosio:2002ex}. In either case, one needs very good control on 
theoretical uncertainties.

As an example, let us consider FCNCs in the down-quark sector, such as
$b \to s$ and $b \to d$ transitions. The dominating short-distance contribution 
within the standard model as well as possible new physics contributions in an
MFV scenario are proportional to the combination $|V_{ts} V_{tb}^*| \, m_t^2$ and
$|V_{td} V_{tb}^*|\, m_t^2$, respectively, and hence the \emph{relative}
strength of the two processes will remain unchanged in MFV.
Still, the analysis may be obscured by the problem of computing the relevant
hadronic matrix elements, for instance the hadronic form factors for
$B \to K^*\gamma$ and $B \to \rho\gamma$ decays \cite{Bosch:2001gv,Ali:2001ez,Beneke:2004dp,Ball:2006nr}. Within the present
uncertainties, the determination of $|V_{td}/V_{ts}|$ from these decays is compatible
with the global CKM fit \cite{CKMfitter}, in particular with the complementary determination
from $\Delta M_{B_d}/\Delta M_{B_s}$. From this we may conclude that new physics
effects in these observables are either absent or supressed via MFV.
Improving the experimental and theoretical errors in both observables in the future 
might reveal a mismatch between the independent determinations of $|V_{td}/V_{ts}|$ 
which would point towards non-minimimal flavour violation.


In the following paragraphs we consider a few more examples, 
where we expect sizeable phenomenological implications within our
particular ansatz for nMFV.

\subsubsection{nMFV: Right handed Spurion $Y_R$}
By construction, the inclusion of an independent spurion $Y_R$
enhances the possible new-physics effects for right-handed transitions,
in particular it directly induces right-handed charged currents.
Therefore, it should be worth looking into right-handed contributions to 
{\em charged}\/
$b \to u$ and $b \to c$ decays, which may be significant despite the
fact that the (left-handed) SM decay is not loop suppressed. In 
particular, the
semileptonic $b \to u$ and $b \to c$ decays will be affected.
The standard methods to extract the SM value for
$|V_{ub}/V_{cb}|$ from exclusive and inclusive decay modes might still
be applicable in MFV scenarios, but in an  nMFV scenario involving the 
right-handed spurion $Y_R$ sizeable pollutions from 
right-handed quarks are expected to alter the result.

This may show up as an 
inconsistency like the
presently observed 1-$\sigma$ tension between $|V_{ub}/V_{cb}|$ and the 
$\sin2\beta$
value from $B \to J/\psi K_s$ within the global fit of the CKM triangle 
\cite{CKMfitter}.
It should also be stressed that nMFV contributions in exclusive and 
inclusive analyses will be rather different, and hence also the present tension between the 
exclusive and inclusive value for $|V_{ub}|$ could be attributed to such an effect.
A strategy to directly test the left-handedness of $b \to c$ transitions from a 
moment analysis of inclusive spectra has recently been discussed in 
\cite{Dassinger:2007pj}.

Other effects of $Y_R$ could show up in channels, in which
the relative enhancement is pronounced by the fact that the SM
contribution is strongly suppressed by the GIM mechanism. A
well-known example is $D^0$-$\bar D^0$ mixing, which is predicted to
proceed very slowly in the SM 
(see the reviews \cite{Burdman:2003rs,Petrov:2006nc} and references therein).
The phenomenological analysis of $D^0-\bar D^0$ mixing is complicated by
the presence of various contributions from different short- and long-distance
scales to the off-diagonal term in the mass matrix
\beq
  2 m_D  \left( M - \frac{i}{2} \Gamma \right)_{12}
&=&
  \langle \bar D^0| {\cal H}_{\rm eff}^{\Delta C=-2} |D^0\rangle
+
  \sum_n \, \frac{\langle \bar D^0| {\cal H}_{\rm eff}^{\Delta C=-1}|n\rangle\langle n| {\cal H}_{\rm eff}^{\Delta C=-1}|D^0\rangle}{M_D - E_n + i\epsilon} \,.
\label{DDbar}
\eeq
In the SM the short-distance contributions in ${\cal H}_{\rm eff}^{\Delta C=-2}$
are dominated by box diagrams with down-type quarks from the first and second
family
\beq
  {\cal H}_{\rm eff}^{\Delta C=-2}& \simeq & \frac{G_F^2}{4\pi^2} \,
  |V_{cs}^* V_{cd}|^2 \, \frac{(m_s^2-m_d^2)^2}{m_c^2} \left( {\cal O} + 2 {\cal O'} \right)
\label{Heff2}
\eeq
where ${\cal O} = [\bar u \gamma_\mu (1-\gamma_5) c]^2$ and ${\cal O'}=
[\bar u (1+\gamma_5) c]^2$. The unitarity of the CKM matrix (neglecting the small
contribution from $V_{ub}$) leads to a double-GIM supression.
The power-counting
w.r.t.\ the Wolfenstein parameter (table \ref{tab:pc}) yields
\beq
  |V_{cs}^* V_{cd}|^2 \, \frac{(\hat m_s^2-\hat m_d^2)^2}{\hat m_c^2} & \sim & \lambda^{14(12)} \,.
\label{ds}
\eeq
The contribution from bottom quarks in the loop, proportional to
$$
  (V_{cb}^* V_{ub})^2 \, \hat m_b^2 \sim \lambda^{14} \,,
$$
is usually neglected.
Notice that the light quarks in the box diagram are off-shell by an amount
of order $m_c^2$ only, which explains the factor $1/m_c^2$ in (\ref{Heff2}) 
and implies that the effective interactions in (\ref{Heff2}) are {\em not}\/ entirely due to
short-distance effects \emph{at the electroweak scale}. 

In contrast, new heavy particles (e.g.\ squarks or non-standard scalars)
could induce $|\Delta C|=2$ transitions at genuinely short-distance scales. 
In MFV the flavour coefficient cannot be larger than $(V_{cb}^* V_{ub})^2 \, \hat m_b^4 \sim \lambda^{18}$, and again we do not expect any sizeable effects. 
In nMFV with spurion $Y_R$, we may, for instance, consider the contribution from purely
right-handed four-quark operators
\beq
  {\cal H}_{\rm eff}^{\Delta C=-2} & \ni & \frac{c_{RR}}{\Lambda_{\rm NP}^2}
  \left( \sum_{D} R_{uD} R_{cD}^* \right)^2 
  \left[\bar u_R \gamma_\mu c_R \right]^2 
\label{Opp}
\eeq
The power counting for the flavour coefficient yields
$\left( \sum_{D} R_{uD} R_{cD}^* \right)^2 = \lambda^{6(4)}$
in the FN scenario (\ref{eq:YRscale}). In the more conservative democratic
scenario (\ref{YRdem}), we obtain $\lambda^{16(12)}$, 
which is close to/the same as in (\ref{ds}). 
In this case, the NP effects might still compete with the SM ones,
if the overall coefficient $c_{RR}$ in (\ref{Opp}) is due to tree-level
processes and not loop-supressed as in the SM.
Similarly, right-handed nMFV operators in ${\cal H}_{\rm eff}^{\Delta C=-1}$
may significantly change the long-distance contributions to $D^0$-$\bar D^0$ mixing
in (\ref{DDbar}).

The short-distance contributions to
$\Delta F = 2$ transitions involving down-type quarks, which are
relevant for  $K^0$-$\bar{K}^0$ and $B^0$-$\bar{B}^0$ mixing, are
dominated by internal top-quark loops in the SM. 
The comparison of SM/MFV
and purley right-handed nMFV contributions to the $|\Delta S|=2$ Hamiltonian reads 
\beq
  \mbox{SM:} &&  \hat m_t^2 \, (V_{ts}^* V_{td})^2 \sim \lambda^{10} \,,
\cr
  \mbox{FN (\ref{eq:YRscale}):} && \left(\sum_U R_{Us}^* R_{Ud}\right)^2 \sim \lambda^{6(2)}
\cr 
  \mbox{democratic (\ref{YRdem}):} && \left(\sum_U R_{Us}^* R_{Ud}\right)^2 \sim \lambda^{16(12)}
\eeq
Similarly, for $|\Delta B|=2$ and $|\Delta S|=0$ one has
\beq
  \mbox{SM:} &&  \hat m_t^2 \, (V_{tb}^* V_{td})^2 \sim \lambda^{6} \,,
\cr
  \mbox{FN (\ref{eq:YRscale}):} && \left(\sum_U R_{Ub}^* R_{Ud} \right)^2 \sim \lambda^{4(2)}
\cr 
  \mbox{democratic (\ref{YRdem}):} && \left( \sum_U R_{Ub}^* R_{Ud} \right)^2 \sim \lambda^{16(12)}
\eeq
and for $|\Delta B|=|\Delta S|=2$
\beq
  \mbox{SM:} &&  \hat m_t^2 \, (V_{tb}^* V_{ts})^2 \sim \lambda^{4} \,,
\cr
  \mbox{FN (\ref{eq:YRscale}):} && \left(\sum_U R_{Ub}^* R_{Us} \right)^2 \sim \lambda^{2(4)}
\cr 
  \mbox{democratic (\ref{YRdem}):} && \left( \sum_U R_{Ub}^* R_{Ud} \right)^2 \sim \lambda^{16(12)}
\eeq
Therefore, the relative effect
of nMFV contributions involving right-handed quarks might be sizeable
in $K^0$-$\bar K^0$ mixing if the power-counting for the matrix $R_{ij}$ 
is close to the (probably unrealistic) FN scenario. 
In all other cases the nMFV effects will in general be less
dramatic than in $D^0$-$\bar{D}^0$ mixing.
On the other hand, the hadronic uncertainties, in particular in the 
case of $B^0$-$\bar{B}^0$ mixing, are under somewhat better control.

\subsubsection{nMFV: Octet Spurions}


We have seen in the above example, that including the nMFV spurion
$Y_R$ we can generate all possible quark-bilinear flavour structures 
with at most {\em two}\/ spurion insertions (in contrast to up to
four in MFV scenarios). Clearly, allowing for the complete set of
nMFV spurions, each flavour transition with a particular chirality
structure has its own spurion. Actually, in the simple FN scenario 
discussed above, the $Z_{L,U,D}$ spurions are allowed and their
power-counting is again fixed by the $a_i$ and $b_i$ quantum numbers. 
As a result of the triangle inequalities, 
the $\bar Q_L Q_L$, $\bar U_R U_R$ and $\bar D_R D_R$ transitions 
can have even larger flavour coefficients than in the $Y_R$ scenario
discussed above.

In cases, in which only left-handed new physics interactions 
appear, the only possible new elementary spurion is $Z_L$.   
An example of such a case is the
Littlest Higgs Model with T-Parity 
\cite{Arkani-Hamed:2002qy,Cheng:2004yc,Low:2004xc}
whose flavour structure has been considered in some detail in 
\cite{Hubisz:2005bd,Blanke:2006sb}.
In these models the left-handed standard fermions couple to left-handed mirror 
fermions via heavy gauge bosons. The flavour structure of these couplings is 
described by two unitary matrices $V_{Hu}$ and $V_{Hd}$, which can be written as
\begin{equation}
  V_{Hu} = V_H^\dagger \, V_{u_L} \,, \qquad
  V_{Hd} = V_H^\dagger \, V_{d_L} \,,
\end{equation}
and satisfy the constraint 
$
V_{Hu}^\dagger V_{Hd} = V_{\rm CKM} \,.
\label{lhrel}
$
At low scales,
the nMFV effects in this model appear due to the mass splitting of the
mirror fermions, such that we may define
\begin{equation}
   V_{H}  \, M_{\rm mf} \, V_{H}^\dagger = \overline{M}_{\rm mf} + V_{H} \, \Delta M_{\rm mf} \, 
V_{H}^\dagger \ \equiv \  \overline{M}_{\rm mf} \left(1 + Z_L \right) \,.
\end{equation}
where $M_{\rm mf}$ is the diagonal mass matrix of the mirror fermions,
$\overline{M}_{\rm mf}$ is their  average mass and $\Delta M_{\rm mf}$
their mass splitting.
If the relative mass splitting $\Delta M_{\rm mf}/\overline{M}_{\rm mf}$ and/or
the off-diagonal matrix elements in $V_{H}$ are sufficiently small,
the littlest Higgs models satisfy our criteria for nMFV.

The minimal super-symmetric extension of the SM introduces new flavour
structures through the soft SUSY-breaking sector. The tri-linear
squark-Higgs couplings transform in the same way as the SM Yukawas.
If one does not allow for generic flavour violation, they
are naturally described in MFV, $A_{ij}^U \propto (Y_U)_{ij}$
and $A_{ij}^D \propto (Y_D)_{ij}$. The squark mass terms transform
as octet spurions, $Z_{L,U,D}$.

Certainly, without a compelling theory of flavour-breaking within a given
new-physics model, it will be extremely difficult to disentangle the effects
of the nMFV spurions $Y_R$, $Z_{L,U,D}$. Nevertheless, we think that
the possibility to classify different flavour-breaking effects beyond MFV 
alone, may be helpful for phenomenological studies which aim to constrain the
flavour sector of physics beyond the SM.

\section{Conclusions}

\label{sec6}
%

In this paper we have proposed a model-independent scheme to classify
new physics contributions to flavour transitions beyond the popular
assumption of minimal flavour violation. In the effective-field-theory approach
to MFV, all flavour transitions can be expressed in terms of fundamental
spurion fields $Y_U$ and $Y_D$ which transform as $(3,\bar 3,1)$ and
$(3,1,\bar 3)$ under the flavour group 
$ SU(3)_{Q_L} \times  SU(3)_{U_R} \times  SU(3)_{D_R}$. In the mass eigenbasis,
$Y_U$ and $Y_D$ are given in terms of quark masses and CKM elements.

We define next-to-minimal flavour violation (nMFV) by allowing new spurion fields 
(\ref{eq:new}), satisfying a particular power-counting in Wolfenstein-$\lambda$
which is constrained by the inequalities~(\ref{ineq}). Depending on the 
considered nMFV spurion and the assumed power-counting, we can enhance certain
flavour decay channels with respect to the SM/MFV. We have worked out the 
specific example of an nMFV spurion $Y_R \sim (1,3,\bar 3)$ which couples to
right-handed quarks. We have found that $Y_R$ can lead to sizeable new-physics 
contributions in neutral $D$\/-meson and kaon decays, as well as in charged 
right-handed $b \to u$ and $b \to c$ transitions.
Our classification scheme may be helpful as a starting point for studies
of flavour violation beyond the SM in the era of the new collider experiments
at the LHC and precision measurements at Super-B factories 
(see \cite{LHCflavor}). 


\section*{Acknowledgements}

This work was supported by the German Research Foundation (DFG)
under contract No.\ MA1187/10-1, and by the German Minister of 
Research (BMBF, contract No.\ 05HT6PSA).


\end{document}